\documentclass[conference, 10 pt]{IEEEtran}
\IEEEoverridecommandlockouts 

\usepackage{graphicx} 
\usepackage{caption} 
\usepackage[utf8]{inputenc}
\usepackage{amsmath}
\usepackage{bm} 
\usepackage{amssymb}
\usepackage{color}
\usepackage{bbm} 
\usepackage{verbatim} 
\usepackage{breqn} 
\usepackage{enumitem} 
\usepackage{amsthm}
\usepackage{amsfonts}
\usepackage{lipsum}

\usepackage{float}

\usepackage{adjustbox}
\usepackage{dsfont}

\def\*#1{\mathbf{#1}}
\newcommand{\vect}[1]{{\mathbf{#1}}}
\newcommand{\mat}[1]{{\mathbf{#1}}}
\renewcommand{\sf}[1]{{\mathsf{#1}}}
\newtheorem{theorem}{Theorem}

\newtheorem{definition}{Definition}

\usepackage{amsmath}

\usepackage{cite}

\usepackage{tikz}
\usepackage{amsmath,amssymb,color,graphicx,caption,subcaption,comment,tikz,url,latexsym} 
\usepackage{algorithmic}
\usepackage{graphicx}
\usepackage{textcomp}
\usepackage{pgfplots}
\usepackage{xcolor}
\usepackage{url}
\usepgfplotslibrary{units}
\usetikzlibrary{spy,backgrounds}
\usepackage{pgfplotstable}

\usepackage{multirow} 

\newcommand{\ab}[1]{{\color{black}#1}}

\begin{document}

\title{Unveiling Covert Semantics: Joint Source-Channel Coding Under a Covertness Constraint}

\author{Abdelaziz Bounhar, Mireille Sarkiss, Michèle Wigger}
    \author{
    \IEEEauthorblockN{Abdelaziz Bounhar$^{\star}$, Mireille Sarkiss$^{\S}$, Mich\`ele Wigger$^{\star}$}
        \IEEEauthorblockA{$^{\star}$LTCI, T\'{e}l\'{e}com Paris, Institut Polytechnique de Paris, 91120 Palaiseau, France
    \\\{abdelaziz.bounhar, michele.wigger\}@telecom-paris.fr}
        \IEEEauthorblockA{$^{\S}$SAMOVAR, T\'{e}l\'{e}com SudParis, Institut Polytechnique de Paris, 91120 Palaiseau, France
	\\\{mireille.sarkiss\}@telecom-sudparis.eu}
}

\maketitle

\allowdisplaybreaks[4]

\begin{abstract}
    The fundamental limit of Semantic Communications (joint source-channel coding) is established when the transmission needs to be kept covert from an external warden.
    We derive information-theoretic achievability and matching converse results and we show that source and channel coding separation holds for this setup.
    Furthermore, we show through an experimental setup that one can train a deep neural network to achieve covert semantic communication for the classification task. Our numerical experiments  confirm our theoretical findings, which indicate that for reliable joint source-channel coding  the number of transmitted source symbols can only scale as the square-root of the number of channel uses. 
\end{abstract}

\begin{IEEEkeywords}
Semantic Communication, Joint Source-Channel Coding, Physical Layer Security, Covert Communication
\end{IEEEkeywords}

\vspace{-0.2cm}
\section{Introduction}
\label{section:introduction}
\IEEEPARstart{S}{}emantic Communication refers to the  emerging communication paradigm where the transmitter sends only the semantics of a source but not the entire source itself  \cite{gunduz_beyond_transmitting_bits}. 
Thanks to the significant gains in bandwidth efficiency, this communication paradigm has attracted great attention in the community and has the potential to be part of next generation communication networks. For traditional communication, 
Shannon's separation theorem \cite{shannon_seperation} implies that without loss in optimality, one can establish semantic communication by simple concatenation of optimal source (compression) and channel codes. 
However, while above optimality only holds in the asymptotic regime of infinite blocklengths, for finite blocklengths a joint design of the source and channel codes can yield improved performances  \cite{verdu_lossy_jscc_finite_blocklength}. 
Indeed, various practical joint source-channel codes have been proposed in the literature, where recent advances particularly focus on implementations using Deep Neural Networks (DNNs)  \cite{deep_jscc_img_generative_gunduz,
deep_secure_jscc_image_gunduz, deep_jscc_img_retrieval_at_edge_gunduz, deep_jscc_txt_1, deep_jscc_txt_2}. 

\ab{Moreover, privacy and security are becoming crucial for communication in many applications, see}
\cite{deep_wiretap_secure_jscc_gunduz, deep_min_leakage_secure_jscc_gunduz, phy_layer_adv_robustness_1}. 
\ab{In this spirit}, 
\cite{deep_secure_jscc_image_gunduz, deep_min_leakage_secure_jscc_gunduz} 
proposed to leverage a DNN-based architecture to simultaneously minimize the distortion of the reconstruction at the legitimate receiver,  while \ab{also} restricting information leakage to potential eavesdroppers.
In this work, we propose a similar \ab{DNN}-based implementation, which however respects the more stringent security constraint that  potential attackers should not only be unable to learn about the transmitted source, but \ab{even} stay agnostic of the \ab{mere} fact that communication is going on. We are thus imposing the constraint that semantic communication be \ab{\emph{Covert}}, i.e., undetectable to external eavesdroppers. 
Our implementation shows that \ab{covert semantic communication} for the \ab{classification} task can be achieved through a DNN architecture, and that  the number of extracted features should be in the order of the square-root of the number of channel uses used for communication. 
This  reminds the well-known square-root law of covert data communication, which was established in \cite{bash_first, bloch_first, ligong_first}.
\ab{Similar observations were noted in studies examining covert detection \cite{covert_detection_isit24} and others involving both covert and non-covert communication \cite{ours_first, covert_mac_icc24}.}

In this work, we \ab{endorse} our numerical experiments with a rigorous information-theoretic analysis that establishes the fundamental limits of joint source-channel coding (JSCC) under a covertness constraint, i.e. the information theoretic limits of  covert semantic communication. 
Our results provide necessary and sufficient conditions under which  covert semantic communication is possible.
These conditions in particular imply that separate source-channel coding is optimal for covert semantic communication, and that the number of source symbols should  scale at most as the square-root of the number of channel uses, similarly as for \ab{covert} data communication. 
Often this square-root scaling is not a problem in semantic communication as the extracted number of features typically occupies a small space. 
The main contribution of our information-theoretic results is the converse proof where we show that \ab{a} separate source-channel coding architecture is optimal in the asymptotic regime of infinite blocklengths.

In brief, this paper makes the following contributions:
\begin{itemize}
    \item We introduce and study the problem of joint source-channel coding under a covertness constraint. 
    \item We show that source-channel separation is optimal in this setup by deriving  matching \ab{information-theoretic} achievability and converse proofs. 
    This establishes necessary and sufficient conditions for the distortions that are achievable in covert joint source-channel coding.
    \item Our experimental setup showcases that a Deep Neural Network can achieve covertness \ab{when transmitting semantic information for a classification task}. 
    The experimental results  confirm our theoretical findings and show that the classification task can only be achieved if the number of extracted features is in the order of the square-root of the number of channel uses. 
\end{itemize}
\section{Information-Theoretic Approach}
\subsection{Notation}
\label{sec:notation}

We follow standard notations in \cite{ours_first, cover, Csiszarbook}. 
In particular, we denote a random variable by $X$ and its realization by $x$.
We write $X^{n}$ and $x^{n}$ for the tuples $(X_1,\ldots, X_n)$ and $(x_1,\ldots, x_n)$, respectively, for any positive integer $n > 0$. 
For a distribution $P$ on $\mathcal{X}$, we note its product distribution on $\mathcal{X}^{n}$ by $P^{\otimes n}(x^{n}) := \prod_{i=1}^{n} P(x_{i})$. 
For two distributions $P$ and $Q$ on $\mathcal{X}$, $\mathbb{D}(P\|Q):=\sum_{x \in \mathcal{X}} P(x)\log(\frac{P(x)}{Q(x)})$ denotes the Kullback-Leibler (KL) divergence between $P$ and $Q$, whilst the chi-squared test is denoted $\chi^2(P\|Q):=\sum_{x \in \mathcal{X}} \frac{(P(x) - Q(x))^2}{Q(x)}$.
Finally, the logarithm function \ab{is} understood in base 2, so the results are in bits.

\subsection{Problem statement}
\label{sec:problem_statement}

\begin{figure}[h]
   \centering
   \includegraphics[scale=1]{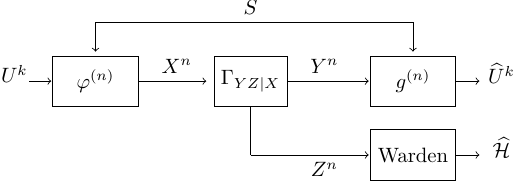}
   \caption{Covert semantic communication system.} 
   \label{fig:setup}
\end{figure}

\label{section:problem_setup_and_main_result}
Consider the \ab{JSCC} problem in Figure \ref{fig:setup}.
The transmitter wishes to communicate a sequence $U^k\in \mathcal{U}^k$ that is drawn i.i.d. according to a given distribution $P_U$ (for a given $k \geq 0$ and an arbitrary finite set $\mathcal{U}$) to a legitimate receiver in the presence of a warden, while tolerating a defined level of distortion $D \geq 0$ at the legitimate receiver. 
The transmitter and the legitimate receiver also share a secret-key $S$, which  is  uniformly distributed over a finite set $\mathcal{S}$ of sufficient large size.\footnote{It has been proved in \cite[Section IV-c]{bloch_first} that generally a \ab{secret-}key size  $\log |\mathcal{S}|$ that scales  as $\sqrt{n}$ suffices to achieve covertness. \ab{When the channel the legitimate receiver is better than the channel to the warden, no secret-key is required at all.} In a similar way, and using the convexity of the Kullback-Leibler divergence\ab{,} one can show that the same  also holds for the setup in this paper. Details are omitted due to lack of space.} 
Communication must remain covert, i.e., an external warden  should not be able to detect the presence of communication.

Technically speaking we have two hypotheses:  under $\mathcal{H}=1$ the communication takes place as described above, while 
under $\mathcal{H}=0$ the transmitter remains silent (sends the "off-symbol"). The transmitter and the legitimate receiver both know $\mathcal{H}$, which however has to remain undetectable to the warden. The receiver and the warden observe channel outputs 
produced by a Discrete and Memoryless Channel (DMC) with a known transition law $\Gamma_{YZ|X}$ and given finite input and \ab{output} alphabets $\mathcal{X}, \mathcal{Y}, \mathcal{Z}$.  That means, if the $X^n=x^n$ then the for any $i$ the $i$-th output symbols $Y_i$ and $Z_i$ observed at the legitimate receiver and the warden are generated from the $i$-th input $x_i$ according to the conditional  laws $\Gamma_{Y \mid X}(\cdot | x_i)$ and $\Gamma_{Z \mid X}(\cdot | x_i)$.\footnote{Notice the generality of this \ab{channel} model that even allows the modeling of fast fading channels.} For simplicity, we assume a binary input alphabet  $\mathcal{X}=\{0,1\}$ and  consider that $0$ is the "off-symbol". We can now describe the communication model and the constraints in full mathematical details.


\textit{\underline{Under $\mathcal{H}=0$:}} the transmitter sends the all-zero sequence 
\begin{align}
X^n&=0^n.
\end{align}

\textit{\underline{Under $\mathcal{H}=1$:}} the transmitter applies some  encoding function $\varphi^{(n)} \colon \mathcal{U}^k \times \mathcal{S} \to \mathcal{X}^n$ to its sequence $U^k$ and sends the resulting codeword 
\begin{equation}
X^n=\varphi^{(n)}(U^k,S)
\end{equation} over the channel. 
For readability\ab{,} we will also write $x^n(u^k,s)$ instead of $\varphi^{(n)}(u^k,s)$.

The legitimate receiver  decodes the desired sequence $U^k$ based on its observed output sequence $Y^n$ and the secret key $S$.
Thus, under $\mathcal{H}=0$ it does nothing whereas under $\mathcal{H}=1$ it uses a decoding function $g^{(n)}\colon \mathcal{Y}^n \times \mathcal{S} \to \mathcal{\widehat{U}}^k$ to produce the guess
\begin{equation}
\widehat{U}^k = g^{(n)}(Y^n, S),
\end{equation}
over a given reconstruction alphabet $\mathcal{\widehat{U}}^k$, which can differ from $\mathcal{U}^k$.
Allowing for a  general reconstruction alphabet $\hat{\mathcal{U}}$ \ab{enables the consideration of} more general reconstruction tasks, such as not reconstructing the entire source symbols but only a feature thereof. (In this case $\hat{\mathcal{U}}$ would be the feature space.)

Decoding performance under $\mathcal{H}=1$ of a \ab{pair} of encoding and decoding functions $(\varphi^{(n)}, g^{(n)})$ is measured by a bounded per-letter distortion measure $d(\cdot, \cdot)$.
We require the average per-block distortion to be less or equal to a given positive threshold $D$,
\begin{IEEEeqnarray}{rCl}
\mathbb{E} \left[\frac{1}{k} \sum_{i=1}^k d(U_i,\widehat{U}_i) \right] \leq D, \label{eq:avg_distortion_def}
\end{IEEEeqnarray}
where expectation is over the random source sequence $U^k$ and the randomness in the channel.
It is assumed that the distortion measure $d(\cdot, \cdot)$ is such that for any $u \in \mathcal{U}$, there exists a reconstruction symbol $\widehat{u} \in \mathcal{\widehat{U}}$ that has zero distortion, i.e. $d(u, \widehat{u}) = 0$.

\noindent Communication is subject to a covertness constraint at the warden, which observes the channel outputs $Z^n$. Under $\mathcal{H}=1$, the warden's output distribution is thus
\begin{IEEEeqnarray}{rCl}
\label{eq:def_Q_C_n}
\widehat{Q}^{n}(z^{n}) &\triangleq& \frac{1}{|\mathcal{S}|}  \sum_{s \in \mathcal{S}} \sum_{u^k} P_U^{\otimes k}(u^k) \Gamma^{\otimes n}_{Z|X} (z^n| x^n(u^k,s)),
\end{IEEEeqnarray}
whereas under $\mathcal{H}=0$ it is
\begin{IEEEeqnarray}{rCl}
\Gamma^{\otimes n}_{Z|X} (z^n| 0^n).
\end{IEEEeqnarray}
Our covertness metric is the KL-divergence between these two output distributions $\mathbb{D}\left(\widehat{Q}^{n} \; \big\| \; \Gamma^{\otimes n}_{Z|X} ( \cdot | 0^n)\right)$.
The choice of this measure is justified by the fact that any test satisfies \cite{book_Testing_Statistical_Hypotheses} $\alpha + \beta \geq 1 - \mathbb{D}\left(\widehat{Q}^{n} \; \big\| \; \Gamma^{\otimes n}_{Z|X} ( \cdot | 0^n)\right)$, for $\alpha$ and $\beta$ denoting the probabilities of miss-detection and false alarm, respectively.
Therefore, ensuring a negligible $\mathbb{D}\left(\widehat{Q}^{n} \; \big\| \; \Gamma^{\otimes n}_{Z|X} ( \cdot | 0^n)\right)$ is \ab{sufficient} to achieve covertness.

Our problem is thus multi-objective in the sense that we not only wish to satisfy the distortion constraint \eqref{eq:avg_distortion_def}, but also a vanishing detectability capability at the warden $\mathbb{D}\left(\widehat{Q}^{n} \; \big\| \; \Gamma^{\otimes n}_{Z|X} ( \cdot | 0^n)\right)$.
These constraints are reflected in the following definition
\begin{definition}
    Let $k=f(n)$ for a given function $f(\cdot)$ on appropriate domains and $\{\delta_n\}_{n\geq \ab{1}}$ be a sequence tending to 0 as the blocklength $n\to \infty$.
    A source-channel pair $(P_U, \Gamma_{YZ \mid X})$ is $(D,\delta_n)$-admissible under a covertness constraint if there \ab{exists} a sequence of encoding and reconstruction functions $\{\varphi^{(n)}, g^{(n)}\}_n$ satisfying the two conditions
    \begin{IEEEeqnarray}{rCl}
         \limsup_{k \to \infty} \mathbb{E} \left[\frac{1}{k} \sum_{i=1}^k d(U_i,\widehat{U}_i) \right] \leq D , \label{eq:distortion_condition_def} \\
        \mathbb{D}\left(\widehat{Q}^{n} \; \big\| \; \Gamma^{\otimes n}_{Z|X} ( \cdot | 0^n)\right) \leq \delta_n, \qquad \forall n. \label{eq:covertness_condition_def}
    \end{IEEEeqnarray}
\end{definition}


\subsection{Information-Theoretic Results}
\label{sec:information_theoretic_results}
In this section, we characterize the fundamental limits of our covert JSCC setup. Our results show that if one wishes to attain a non-trivial distortion $D$ and at the same time  satisfy a  covertness constraint $\delta_n$ (see \eqref{eq:covertness_condition_def}) then the number of source symbols $k$ can scale at most \ab{proportional to} $ \sqrt{ n \delta_n}$ (see Result 2). We \ab{say that a distortion is} trivial if it  can be achieved without communication by having  the receiver produce a constant reconstruction symbol, i.e., if 
\begin{equation}
D\geq D_{\textnormal{trivial}}:= \min_{\widehat{u} \in \mathcal{\widehat{U}}} \mathbb{E}[d(U,\widehat{u})].
\end{equation}   If the number of source symbols $k$ scales as $\sqrt{n \delta_n}$,  we show that a distortion is achievable if, and only if,  it is achievable by a separate source-channel code, i.e.,  the number of symbols required for compressing the source is less than the channel capacity multiplied by the bandwidth mismatch factor (Result 3).
Finally, (in Result 1) we also show that if the number of source symbols scales slower than $\sqrt{n \delta_n}$, then arbitrary small  distortion levels $D\geq 0$ can be  achieved.

Before stating our main result, recall the definition of the standard
 rate-distortion function \cite{cover}
\begin{IEEEeqnarray}{rCl}
R(D) & \triangleq & \min_{P_{\widehat{U} \mid U}(\widehat{u} \mid u) : \mathbb{E}_{P_UP_{\widehat{U} \mid U}} \left[ d(\widehat{U}, U) \right] \leq D } \mathbb{I}(\widehat{U}, U), 
\end{IEEEeqnarray}
and of the covert capacity \cite{bloch_first, ligong_first} \ab{for binary input alphabets}
\begin{IEEEeqnarray}{rCl}
   C_{\textnormal{covert}} &\triangleq & \sqrt{2}  \frac{ \mathbb{D} \left( \Gamma_{Y \mid X}(\cdot \mid 1) \| \Gamma_{Y \mid X}(\cdot \mid 0) \right)} {\sqrt{ \chi^2(\Gamma_{Z \mid X}(\cdot \mid 1) \| \Gamma_{Z \mid X}(\cdot \mid 0))}}.  \label{eq:covert_capacity}
\end{IEEEeqnarray}


Recall that the number of source symbols $k=f(n)$ for a given function $f(\cdot)$.

\begin{theorem}
    \label{main_theorem}
 For any given function $f(\cdot)$ and  vanishing sequence  $\{\delta_n \}_{n\geq 1}$ the following holds. 
    \begin{enumerate}
        \item If
        \begin{equation}
            \lim_{n \to \infty} \frac{f(n)}{\sqrt{n \delta_n}} = 0,\label{eq:zero}
        \end{equation}
        then all nonnegative distortions $D \geq 0$ with finite $R(D)$ are $(D, \delta_n)$-admissible. 
       \item If
        \begin{equation}
            \lim_{n \to \infty} \frac{f(n)}{\sqrt{n \delta_n}} = \infty,\label{eq:infty}
        \end{equation}
        then only  trivial distortions $D \geq D_{\textnormal{trivial}}$ are $(D, \delta_n)$-admissible.        \item If
        \begin{equation}
            \lim_{n \to \infty} \frac{f(n)}{\sqrt{n \delta_n}} = \frac{1}{\gamma}, \label{eq:value_k_theorem_1}
        \end{equation}
        for some $\gamma > 0$, then $D$ is $(D, \delta_n)$-admissible if, and only if,
        \begin{IEEEeqnarray}{rCl}
            R (D) &\leq&  \gamma    C_{\textnormal{covert}} .  \label{eq:th_rate_upper_bound}
        \end{IEEEeqnarray}
    \end{enumerate}

Notice that the parameter $\gamma$ plays the same role as the bandwidth mismatch factor in traditional JSCC. 
\end{theorem}
\begin{IEEEproof}
Result 1) only requires a proof of  achievability and Result 2) only a proof of  converse. Result 3) requires both proofs. The two converses are proved in Appendix~\ref{sec:converse_proof}. We now prove the two achievability results  based on 
the separate source-channel coding architecture in Figure~\ref{fig:source_channel_coding}.
\begin{figure}[h!]
    \centering
    \includegraphics[width=0.49\textwidth]{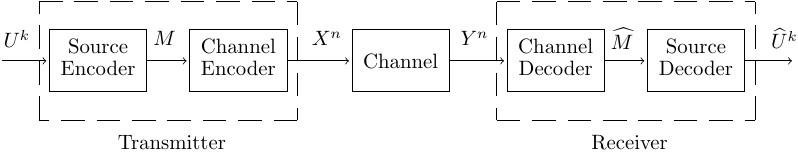}
    \caption{Separate source and channel coding architecture.}
    \label{fig:source_channel_coding}
\end{figure}

 Specifically, the transmitter initially compresses the source sequence $U^k$ into an index $M\in\{1,\ldots, 2^{kR}\}$, which is subsequently encoded into a codeword $X^n$ by a channel code, and then transmitted over the DMC.
A channel decoder observe $Y^n$, a noisy version of $X^n$, and maps it to a  guess $\ab{\widehat{M}}$ of the  index $M$. This index is then 
 used by the source decoder to produce the reconstruction of the source $\widehat{U}^k$.  
It is possible to choose a good lossy compression scheme, such as the likelihood lossy compression scheme in \cite{song_cuff_likelihood_encoder}, so that  the reconstruction $\ab{\widehat{U}}^k$ satisfies the distortion constraint \eqref{eq:distortion_condition_def} whenever $\ab{\widehat{M}}=M$ and the rate $R >R(D)$. On the other hand, it can be shown that for any vanishing sequence $\delta_n$ there exists a good covert channel code     \cite{bloch_first} that conveys the message $M$ with vanishing probability of error and at the same time respects the covertness constraint \eqref{eq:covertness_condition_def} whenever 
\begin{equation}\label{eq:covert_cond}
\lim_{n\to \infty} \frac{k R}{\sqrt{n \delta} }< C_{\textnormal{covert}} .
\end{equation}
Recall that $k=f(n)$ and notice that under Condition~\eqref{eq:zero}, Inequality  \eqref{eq:covert_cond} is satisfied for all finite values of $R$. This  establishes Result 1). Under Condition~\eqref{eq:value_k_theorem_1}, it is possible to find a finite value of $R>R(D)$ satisfying \eqref{eq:covert_cond} whenever 
\begin{equation}
R(D) < \gamma C_{\textnormal{covert}}, 
\end{equation}
thus proving achievability of Result 3). 
This concludes the desired proofs. 
\end{IEEEproof}
\section{Covert Semantic Extraction: Neural Network Experiment}

\begin{figure}[h!]
    \centering
    \includegraphics[width=0.49\textwidth]{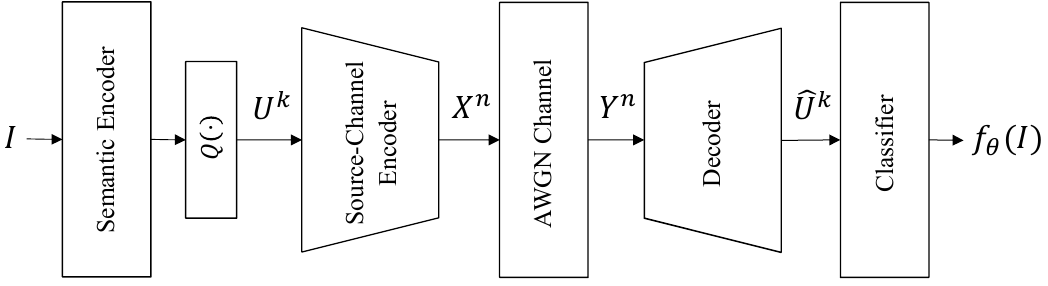}
    \caption{Neural Network architecture for distributed classification  under a covertness constraint.}
    \label{fig:neural_net}
\end{figure}

To illustrate our theoretical findings in a practical context, we train two DNNs for the task of image classification over \ab{an Additive White Gaussian Noise (AWGN)} channel,  see  Figure \ref{fig:neural_net}. 
Specifically, we train a first DNN based on a training \ab{dataset} $\mathcal{D}_{\textnormal{train}}$ of images \ab{and corresponding labels} to implement \ab{the} pair of \ab{Semantic} \ab{Encoder} (feature extractor) and \ab{Classifier, see Figure \ref{fig:neural_net}}. 
We \ab{subsequently} train a second DNN \ab{(on the same training dataset $\mathcal{D}_{\textnormal{train}}$)} to implement the Source-Channel Encoder and Decoder over \ab{the} AWGN channel based on the output sequence $U^k$ produced by the previously trained Semantic Encoder,  so that communication remains undetectable to an external warden.
In contrast to the standard JSCC model, in our experiment the sequence of extracted features $U^k$ (also called semantic vector) is not i.i.d. but has a given distribution \ab{dictated by the Semantic Encoder}.

Our \ab{datasets} $\mathcal{D}_{\textnormal{training}}$ and $\mathcal{D}_{\textnormal{test}}$ both consist of image-\ab{label} pairs $(I, y)$, where $I \in \mathbb{R}^{h \times w \times c}$  (for $h$, $w$ and $c$  the height, width and the number of channels of the image respectively) and $y \in \{1, \ldots, C\}$ (for $C$ indicating the number of classes). 
If we denote by $f_{\theta}(I)$ the output of \ab{the combined two DNNs (\ab{see} Figure~\ref{fig:neural_net}}) when the input image is $I$ and the DNNs parameters fixed after the training are described by $\theta$, then the accuracy on the test data\ab{, which measures the performance of our model on unseen images,} is defined as
\begin{equation}
    \label{eq:accuracy}
\textnormal{Acc}(\theta, \mathcal{D}_{\textnormal{test}}  ) \ab{\triangleq}  \frac{1}{|\mathcal{D}_{\textnormal{test}}|} \sum_{(I,y) \in \mathcal{D}_{\textnormal{test}}} \mathds{1} \{ f_{\theta}(I) = y\}.
\end{equation}

Notice that when combining the two DNNs, we can tweak the parameter $k$ indicating the number of features  to be extracted from the images or equivalently  the number of source symbols that have to be sent over the AWGN channel. 
Based on the theoretical findings in the previous Section \ab{\ref{sec:information_theoretic_results}}, our goal is to show that in order to achieve a satisfactory accuracy, the length $k$ of the semantic vector $U^k$ that \ab{contains the semantic information used} to classify \ab{an} image must \ab{approximately be} $\sqrt{n \delta_n}$ for a small $\delta_n$ and a large blocklength $n$. 
Failure to adhere to this scaling should result in a bad classification accuracy because we expect the channel coding to introduce too many errors. 

\subsection{DNN Architectures and Training}

\ab{As aforementioned,} our first DNN comprises a \emph{Semantic Encoder} and a \emph{Classifier}. The Semantic Encoder maps the input image $\ab{I}$ to a hidden representation \ab{(a vector in $\mathbb{R}^k$)} to which we apply the binary quantization function $Q(x) \triangleq \mathds{1}\{ x > 0 \}$ componentwise.
The output of this quantization procedure then provides the binary \ab{semantic} vector $U^k$ \ab{with} $k$ a parameter that we can choose in our implementation. 
This first DNN is trained so as to minimize the cross-entropy\ab{\footnote{\ab{It can also be viewed as a minimization of the KL-divergence between the true distribution $P$ and the DNN distribution $Q$, i.e. $\mathcal{L}^{\ab{(1)}} = - \lambda_{ce} \cdot \left[ \mathbb{H}(P) + \mathbb{D}(P \| Q) \right] $.}}}, i.e.,  the loss 
\begin{equation}
    \mathcal{L}^{\ab{(1)}} = - \lambda_{ce} \sum_{y=1}^C y \log(p_y),
\end{equation}
\ab{where $p_y$ represents the probability\ab{\footnote{\ab{Our DNN provides probabilities on the classes, which here we denote $\{p_y\}_{y=1}^C$ before deciding on its argmax class.}}} that the DNN gives to the class $y$ whereas $\lambda_{ce}$ is a scaling constant that can be adjusted through tuning.}

The second DNN  implements the Source-Channel Encoder and the Decoder. 
It \ab{simultaneously} seeks to minimize the Hamming distortion error for the reconstruction of the semantic vector $U^k$ \ab{and} aims to achieve covert communication. Instead of constraining the warden's divergence\ab{\footnote{One could \ab{consider} the KL divergence of the histograms by resorting to soft histograms \cite{soft_histogram}. In practice, we find that this leads to poor performance.}} $\mathbb{D}\left(\widehat{Q}^{n} \; \big\| \; \Gamma^{\otimes n}_{Z|X} ( \cdot | 0^n)\right)$, as we did in the previous section, here we attempt to constrain the  power of the  transmit signal $X^n$. 
In some sense, this can be viewed as a more universal approach as the covertness constraint does not rely on a specific model for the warden.  Specifically, we add the loss term
\begin{equation}
    \mathcal{L}_{\textnormal{covert}} = \left| \left( \frac{\frac{1}{n} \sum_{i=1}^n X_i}{\sqrt{n \epsilon}} \right)^2 - 1\right|,
\end{equation}
which is motivated by the Central Limit Theorem which states that the noise uncertainty at a receiver suffering from Gaussian noise is in the order of $\sqrt{n}$. The $\sqrt{\epsilon}$ factor allows to adapt to the desired level of covertness. The  objective when training the second DNN is thus to minimize the loss
\begin{equation}
    \mathcal{L}^{\ab{(2)}} = \lambda_{d} \cdot \frac{1}{k} \sum_{i=1}^k |\widehat{U}_i - U_i| + \lambda_{\textnormal{covert}} \cdot \mathcal{L}_{\textnormal{covert}},
\end{equation}
where $\lambda_{d}$ and $\lambda_{\textnormal{covert}}$ are scaling constants that can be adjusted through tuning.

Notice that the quantizer $Q(\cdot)$ as defined \ab{earlier in this subsection} is non-differentiable and is thus not trainable.
To enable an end-to-end differentiable approach, we resort to the  "Straight-Through Estimator" \cite{ste_paper}, which basically sets the gradients with respect to \ab{the quantizer} to 1 in the backward pass. In fact, the gradients are "straight-through" from the loss function to the model parameters, despite the discontinuity introduced by the discrete sampling in the forward pass.

\subsection{Numerical Results}
We use the MNIST-digit dataset \cite{mnist_paper} with  $\ab{|}\mathcal{D}_{\textnormal{training}}|=50000$ training samples,  $|\mathcal{D}_{\textnormal{test}}|=10000$ test samples, $C=10$ \ab{classes, and $c=1$ channel (black and white images)}.
We train the first DNN within 60 epochs and with a fixed learning rate of $0.01$, while we use $\lambda_{ce}=1$. 
The second DNN is trained with 60 \ab{epochs}, a learning rate of $0.005$, $\lambda_{d} = 10$, $\lambda_{\textnormal{covert}}=10$, $\epsilon = 0.01$, $\delta_n = 0.02$ and the \ab{noise power of the AWGN channel is fixed at 0.63}. 
\ab{For the two DNNs, we set} the batch size to 128 and we use the Adam optimizer \cite{adam} with $\beta =(0.9, 0.999)$ and $\epsilon=10^{-8}$.

We consider two different values for the blocklength $n \in \{512,2048\}$, and start by considering a model where the size of the semantic \ab{vector} $k$ is in the order of $\sqrt{n \delta_n}$, as indicated by Theorem \ref{main_theorem}.

\underline{\textit{1) Square-root covert model}}: 
For $n=512$ we let $k$ take value in $\mathcal{K}_{\textnormal{square-root}}^{(512)} \triangleq \{1,3,4,6,7\}$ and for $n=2048$  we let $k$ \ab{in} $\mathcal{K}_{\textnormal{square-root}}^{(2048)} \triangleq \{2, 4, 5, 8, 10, 11, 12, 14\}$. 
For each value of $n$ we then optimize the accuracy $\textnormal{Acc}(\theta, \mathcal{D}_{\textnormal{test}} )$ \ab{in \eqref{eq:accuracy}} over the value of $k$, and denote the optimal value by $k^*$.
As indicated by Table \ref{table:results_sqrt_covert_model}, the accuracy  increases with larger blocklength $n$ since a larger blocklength provides more room for error correction over the AWGN \ab{channel}.
\ab{Moreover, the number of classes in our experiment is $C=10$ and thus smaller than $2^{k^*}$, so it is beneficial to extract a larger semantic vector than available \ab{classes}.}
\begin{table}[h!]
    \centering
    \begin{tabular}{|c|c|c|c|}
        \hline
        \textbf{Blocklength} & \textbf{Accuracy} & \textbf{Optimal $k^*$} \\ \hline
        512 & 58.45 & 6 \\ \hline
        2048 & 87.44 & 11 \\ \hline
    \end{tabular}
    \caption{Performance under the square-root covert model.}
    \label{table:results_sqrt_covert_model}
\end{table}

Our results  indicate that the training of the DNNs was successful. In particular the joint source-channel   transmission of the semantic vector seems to have been successful, when the  size of the semantic vector $k$  is close to $\ab{2 \cdot}\sqrt{n \delta_n}$, which for the chosen parameters evaluates to $\ab{6.4}$ and $\ab{12.8}$. 

In the following, we further investigate  above  conclusions. 
To this end, we run two additional related models with  sizes of the semantic vector that are in the order of $n$. 
In \ab{particular, for the  Linear covert model 2)} we keep the covertness constraint, 
which we then remove for \ab{the  Linear non-covert model 3)}. 
\ab{The goal is to see whether extracting larger semantic vectors yields better classification performance (i.e. higher accuracy), and whether the conclusions depend  on the imposed covertness constraint.}

\noindent \underline{\textit{2) Linear covert model}}: Here, for $n=512$ the parameter $k$ takes value in 
$\mathcal{K}_{\textnormal{linear}}^{(512)} \triangleq \{102, 409, 512\}$ and for $n=2048$ it takes value in $\mathcal{K}_{\textnormal{linear}}^{(2048)} \triangleq \{409, 1638, 2048\}$. We again optimize over the proposed set of $k$-values. \begin{table}[h!]
    \centering
    \begin{tabular}{|c|c|c|c|}
        \hline
        \textbf{Blocklength} & \textbf{Accuracy} & \textbf{Optimal $k^*$} \\ \hline
        512 & 10.10 & 409 \\ \hline
        2048 & 11.34 & 1638 \\ \hline
    \end{tabular}
    \caption{Performance under the linear covert model}
    \label{table:results_linear_covert_model}
\end{table}

\noindent \underline{\textit{3) Non-covert model}}: Same as the linear covert model, but the loss related to covertness is ignored, i.e., $\lambda_{\textnormal{covert}}=0$.
\begin{table}[h!]
    \centering
    \begin{adjustbox}{margin=5} 
        \begin{tabular}{|c|c|c|c|}
            \hline
            \textbf{Blocklength} & \textbf{Accuracy} & \textbf{Optimal $k^*$} \\ \hline
            512 & 98.88 & 102 \\ \hline
            2048 & 98.95 & 409 \\ \hline
        \end{tabular}
    \end{adjustbox} 
    \caption{Performance under the linear non-covert model}
    \label{table:results_non_covert_model}
    \vspace{-0.5cm}
\end{table}

As  Tables \ref{table:results_linear_covert_model} and \ref{table:results_non_covert_model} show,  the accuracy again increases with the blocklength $n$. However, while under the covertness constraint the achieved accuracies \ab{with the Linear covert model} fall short compared to \ab{the ones with the Square-root covert model}, without the covertness constraint the accuracy is high even at short blocklengths. 
This last  finding indicates that  large semantic vectors are beneficial \ab{to increase accuracy}. 
In contrast,  the low accuracy in the Linear covert model indicates that under a covertness constraint the  probability of error over the communication channel is high when the feature vector is in the order of the blocklength, \ab{thus compromising the overall performance of the classifier}. 

This is perfectly in line with the theoretical findings of our previous Section \ab{\ref{sec:information_theoretic_results}}. 
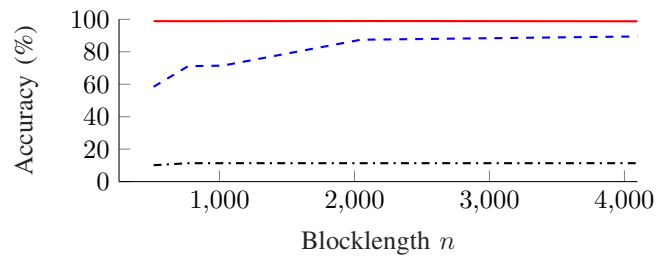
\begin{figure}[!h]
    \centering
    \begin{tikzpicture}
        \begin{axis}[
            width=0.38\textwidth,
            height=0.85in,
            at={(0in,0in)},
            scale only axis,
            xmin=256,
            xmax=4096,
            xlabel style={font=\color{white!15!black}},
            xlabel={Blocklength $n$},
            ymin=0,
            ymax=100,
            ylabel style={font=\color{white!15!black}},
            ylabel={Accuracy (\%)},
            axis background/.style={fill=white},
            title style={font=\bfseries},
            axis x line*=bottom,
            axis y line*=left,
            legend style={legend cell align=left, align=left, draw=white!15!black}
        ]

            

            \addplot [thick, color=black, dashdotted] 
              table[row sep=crcr]{%
            512	10.1\\
            768	11.35\\
            1024	11.35\\
            2048	11.35\\
            4096	11.35\\
            };


    
            \addplot [thick, color=blue, dashed]
              table[row sep=crcr]{%
            512	58.45 \\
            768	71.15 \\
            1024 71.48 \\
            2048 87.44 \\
            4096 89.45 \\
            };


            

            \addplot [thick, color=red, solid]
              table[row sep=crcr]{%
            512	98.88\\
            768	98.86\\
            1024 98.87\\
            2048 98.95\\
            4096 98.81\\
            };

        \end{axis}
        
        \end{tikzpicture}%
        \caption{Accuracy in (\%) as a function of the blocklength $n$ for the three models at SNR=1dB. The solid red curve denotes the Non-covert model, the dashed blue curve the Square-root covert model and the dash-dotted black curve the Linear covert model.}
    
        \label{fig:neural_net_simulation}
\end{figure}

In Figure \ref{fig:neural_net_simulation}, we illustrate the comparison of the accuracies of the three models in function of the  blocklengths $n$. Clearly, the linear covert model saturates and the high probability of communication error severely limits the classification task.
\section{Summary and Discussion}
\label{sec:conclusion}
We established the fundamental limits of semantic communication under a covertness constraint by providing sufficient and necessary conditions for a source to be \ab{reconstructed} with desired distortion at \ab{a} distant receiver. 
In particular, we \ab{have demonstrated the} optimality of source-channel separation for joint source-channel coding under a covertness constraint. 
Moreover, our experimental setup underscores the feasibility of training a deep neural network to accomplish covert semantic communication as long as it suffices to extract a feature vector of length approximately equal to the square-root of the communication blocklength. 
This confirms our theoretical findings \ab{showing} the necessity of the described scaling, \ab{similarly} to the case of covert data communication. 

Future interesting research directions include \ab{extensions to} setups with many users and with power and resource allocation strategies, as well as identifying the minimum secret-key that is required to ensure covertness.
\begin{appendices}
\section{Information-Theoretic Converse Proof}
\label{sec:converse_proof}
Fix a function $f(\cdot)$ and a vanishing sequence $\{\delta_n\}_{\ab{n \geq 1}}$. Consider then a  sequence (one for each $n$) of \ab{JSCC} schemes that \ab{satisfies} both \eqref{eq:distortion_condition_def} and \eqref{eq:covertness_condition_def}. 

Following the same steps as in \cite{bloch_first}, it can be shown that: 
\begin{IEEEeqnarray}{rCl}
\delta_n&\geq& n \frac{\bar{\alpha}_n^2}{2} \cdot \left[\chi^2(\Gamma_{Z \mid X}(\cdot| 1) \| \Gamma_{Z \mid X}(\cdot| 0)) + o(1) \right] \ab{,} \label{eq:converse_upper_bound_divergence_final}
\end{IEEEeqnarray}
where \ab{$o(1)$ is a decreasing function in $n$ and}
\begin{equation}
    \bar{\alpha}_n \triangleq \frac{1}{n} \sum_{i=1}^n \alpha_{n,i}.
\end{equation}

Define $T$ to be uniform over $\{1,\ldots, n\}$ independent of all inputs, outputs, source and reconstruction symbols.  Then: 
\begin{IEEEeqnarray}{rCl}
\mathbb{E} \left[ d(U_T,\widehat{U}_T) \right]  & = &
\mathbb{E} \left[\frac{1}{k} \sum_{i=1}^k d(U_i,\widehat{U}_i) \right]  
 \leq D,\label{eq:d}
\end{IEEEeqnarray}
where the last step holds by  \eqref{eq:distortion_condition_def}. Continue to bound: 
\begin{IEEEeqnarray}{rCl}
	\mathbb{I}(U^k; \widehat{U}^k)
    &=& \mathbb{H}(U^k) - \mathbb{H}(U^k \mid \widehat{U}^k) \\
	&\overset{(a)}{\geq}& \sum_{i=1}^k \mathbb{H}(U_i) - \mathbb{H}(U_i \mid \widehat{U}_i) = \sum_{i=1}^k  \mathbb{I}(U_i; \widehat{U}_i) \IEEEeqnarraynumspace \\
	&=& k \mathbb{I}(U_T; \widehat{U}_T \mid T) 
	\overset{(b)}{\geq} k \mathbb{I}(U_T; \widehat{U}_T)
	\overset{(c)}{\geq}  k R(D), \label{eq:step_1_lb_mutual_info_converse} \IEEEeqnarraynumspace
\end{IEEEeqnarray}
where (a) holds because conditioning reduces entropy and  $U_i$ is independent of  $S$;  (b) holds by the  independence of $U_T$  and  $T$; and $(c)$ holds by the definition of $R(D)$ and because $U_T$ is distributed according to $P_U$ and  $\ab{\widehat{U}}_T$ satisfies \eqref{eq:d}.

Next, notice that the Markov Chain
	$U^k \leftrightarrow (X^n, S) \leftrightarrow (Y^n, S) \leftrightarrow \widehat{U}^k$
implies by the Data Processing Inequality:  
\begin{IEEEeqnarray}{rCl}
\mathbb{I}(U^k; \widehat{U}^k)=
    &\leq& \mathbb{I}(U^k; \widehat{U}^k, S) \overset{(a)}{=} \mathbb{I}(U^k; \widehat{U}^k \mid S) \\
    &\leq& \mathbb{I}(X^n; Y^n \mid S) \\
	&=& \sum_{i=1}^n \mathbb{H}(Y_i \mid Y^{i-1}) - \mathbb{H}(Y_i \mid Y^{i-1}, X^n, S)\IEEEeqnarraynumspace \\
    &\overset{(b)}{\leq}& \sum_{i=1}^n \mathbb{H}(Y_i) - \mathbb{H}(Y_i \mid X_i) \\
    &\overset{(c)}{\leq}& n \bar{\alpha}_n \mathbb{D} \left(\Gamma_{Y \mid X}(\cdot \mid 1) \| \Gamma_{Y \mid X}(\cdot \mid 0) \right)\\
    & \overset{(d)}{\leq } &  \sqrt{n \delta_n}C_{\textnormal{covert}},\label{eq:step_2_ub_mutual_info_converse}
\end{IEEEeqnarray}
where 
$(a)$ holds because $\mathbb{I}(U^k; S)=0$; $(b)$ by the memoryless channel; $(c)$ by \cite[Lemma 1]{bloch_first};  and $(d)$ 
  by \eqref{eq:converse_upper_bound_divergence_final} and  the definition of $C_{\textnormal{covert}}$ in \eqref{eq:covert_capacity}.
\ab{Combining} \eqref{eq:step_1_lb_mutual_info_converse} with \eqref{eq:step_2_ub_mutual_info_converse} \ab{yields} 
\begin{equation}
R(D) \leq \varlimsup_{n \to \infty} \frac{  \sqrt{  n    \delta_n} }{k}  C_{\textnormal{covert}}. \label{eq:iconverse}
\end{equation}
Recalling that $k=f(n)$, we can conclude that whenever 
$\lim_{n\to \infty} \frac{ f(n)}{\ab{\sqrt{n \delta_n}}} =0$, then
Inequality \eqref{eq:iconverse} implies that $R(D)=0$, allowing only for the trivial distortion $D_{\textnormal{trivial}}$, thus establishing the converse to Result 2). 
On the other hand, if 
$\lim_{n\to \infty} \frac{ f(n)}{\ab{\sqrt{n \delta_n}}} =\frac{1}{\gamma}$,
then Inequality \ab{\eqref{eq:iconverse}} is only satisfied if $R(D) \leq \gamma C_{\textnormal{covert}}$, thus establishing the converse to Result 3). 

\end{appendices}

\bibliographystyle{IEEEtran} 
\bibliography{references}  

\end{document}